\documentclass[journal]{IEEEtran}
\ifCLASSINFOpdf
   \usepackage[pdftex]{graphicx}
\else
\fi
\usepackage{amsthm,amssymb,mathtools}
\usepackage{algorithmic}

\usepackage{array}

\ifCLASSOPTIONcompsoc
  \usepackage[caption=false,font=normalsize,labelfont=sf,textfont=sf]{subfig}
\else
  \usepackage[caption=false,font=footnotesize]{subfig}
\fi
\usepackage{fixltx2e}

\usepackage{stfloats}

\usepackage {tikz}
\usetikzlibrary {positioning,shapes,arrows,chains}

\tikzstyle{block} = [draw, fill=black!20, rectangle, 
minimum height=1.5em, minimum width=4em]
\tikzstyle{sum} = [draw, fill=black!20, circle, node distance=3 cm]
\tikzstyle{input} = [coordinate]
\tikzstyle{output} = [coordinate]
\tikzstyle{pinstyle} = [pin edge={to-,thin,black}]

\DeclareMathOperator{\rank}{rank}
\DeclareMathOperator{\tr}{tr}

\newtheorem{thm}{Theorem}

\newtheorem{prop}{Proposition}

\theoremstyle{definition}

\begin{document}
%
\title{Nonlinear Adaptive Algorithms \\on Rank-One Tensor Models}
%
%
%

\author{Felipe C. Pinheiro,
        Cassio G. Lopes,~\IEEEmembership{Senior Member,~IEEE.}}
\maketitle

\begin{abstract}
This work proposes a low complexity nonlinearity model and develops adaptive algorithms over it. The model is based on the decomposable---or rank-one, in tensor language---Volterra kernels. It may also be described as a product of FIR filters, which explains its low-complexity. The rank-one model is also interesting because it comes from a well-posed problem in approximation theory. The paper uses such model in an estimation theory context to develop an exact gradient-type algorithm, from which adaptive algorithms such as the least mean squares (LMS) filter and its data-reuse version---the TRUE-LMS---are derived. Stability and convergence issues are addressed. The algorithms are then tested in simulations, which show its good performance when compared to other nonlinear processing algorithms in the literature.
\end{abstract}


%
\IEEEpeerreviewmaketitle

\section{Introduction}
 
\IEEEPARstart{N}{onlinear} signal processing has its uses whenever the performance of linear techniques start to become inadequate. A few situations that may ask for their use are echo cancellation \cite{echo}, equalization of communication systems \cite{modulation}, acoustics \cite{audio}, or broadband noise canceling \cite{broadband}. 

While an important part of current signal processing research, an usual feature of nonlinear techniques is their high computational complexity, which sometimes render them unusable for some applications, such as those that use slower computers or require low power consumption. This complexity may be high even for modern computers, as is the case of the Volterra series \cite{poly,bk:poly} and its exponentially increasing complexity. 

The Volterra series is an important model because it is able to work with any continuous nonlinearity. It is also a very simple structure that can be described in a linear-in-the-parameters model. Given these advantages, various structures try to work with this model. But many times some restriction is necessary to allow for fast computation. Examples of these approaches are the filters with truncated diagonals \cite{svolt, svolt2}, which works with the central coefficients of the model, which are usually the most significant ones. Another kind of model is based in interpolation techniques \cite{batista1,batista2}, in a way that the algorithms work with only a few coefficients while interpolating the others.

Another aspect of nonlinear signal processing is the difficulty in their analyses, specially if the models of nonlinearity are nonlinear in the parameters. This is the case with cascade structures \cite{cascade}. Some problem of this kind is the lack of global asymptotic stability, with one example shown in this paper. Other models that have yet to be fully analyzed. An example is  ones based in tensor decomposition\cite{favier}.

Tensors are objects that can be accessed via many number of indexes. Vectors are one-index tensors, while matrices are two-index ones. Tensors are also a bridge between the multilinear \cite{greub} world---of multilinear functions, or functions of many variables that are linear in each one of them---and the linear world. As it will be shown in this paper, this property creates a conceptual link between the Volterra series and tensors, with these objects being the natural way of representing a Volterra kernel. Not surprisingly, tensors have indeed been used in nonlinear signal processing-related problem for quite some time \cite{high, volten}.

The tensor representation, and specially tensor rank decomposition \cite{tensors} allows for a dramatic decrease in representational complexity of the model and, should this be exploited in the Volterra series, many low-complexity algorithms may be obtained. Of particular importance is the rank-one approximation \cite{rankone}. This concept has seem some success in approximating linear responses \cite{rupp1,rupp2} with certain structures, and the possible gains to nonlinear signal processing are even more pronounced---exponential reductions in complexity can be achieved \cite{eusipco2016}. The other theoretical advantage of rank-one approximations is their well-posedness, which may not be the case for general low-rank decompositions \cite{illposed}.

This paper focuses on the development of the rank-one approximation of the Volterra series and on the results adaptive algorithms one can derive from the model. For this objective, the paper shall be organized as follows:

Therefore, the paper structure is as follows:
\begin{enumerate}
\item Introduction and notation.
\item A presentation on tensors and rank-one approximations.
\item The Volterra series and its relation to tensors. The decomposable (rank-one) model as a product of FIR linear filters, with low computational complexity.
\item Estimation theory of theory of the decomposable model.
\item The steepest descent algorithm as an iterative solution to the estimation problem.
\item LMS and TRUE-LMS algorithms as instantaneous approximations to the steepest descent. Stability issues and choice of parameters.
\item Simulations: testing performance, stability and applicability.
\item The presence of chaotic behavior.
\item Concluding remarks.
\end{enumerate}

\subsection{Notation}

The notation of the paper follows the one on \cite{bk:sayed}, while introducing new notation when necessary. The conventions are summarized as
\begin{itemize}
\item Scalars and vectors are represented by lowercase letters. (E.g. $x$ and $a$.)
\item Time varying vectors are indexed as $x_i$, while time varying scalars are presented as $a(i)$. The time variable is always $i$ or $j$.
\item Matrices and constants are represented as uppercase letters. (E.g. $R$ and $A$.)
\item Tensors are represented as calligraphic letters. (E.g. $\mathcal{W}$ and $\mathcal{X}$.)
\item Time-varying matrices and tensors follow the vector convention.
\item Tensors by nature have a matrix representation. When interpreted as matrices, the involved operations are to be interpreted as the matrix equivalents. (E.g. the tensor product $\otimes$ is to be interpreted as the Kronecker product \cite{kron}.)
\item This paper uses the classical convention of lower and upper indexes. Column vectors are indexed with an upper index (e.g. $(v)^i$), row-vectors with a lower one (e.g. $(w)_j$) and matrices with both an upper one and a lower one (e.g. $(A)^i_j$). The convention for general tensors will be explained in through the paper.
\end{itemize}

\section{Tensors}
Just like vectors are objects parameterized by a single index and matrices by two, tensors are objects parameterized by an arbitrary number $K$ of indexes---such number being called the order of the tensor. For example, an order $K$ tensor $\mathcal{T}$ may be indexed as $(\mathcal{T})^{i_1\dots i_K}$. Such notions have been used by some developments in signal processing, usually dealing with multi-dimensional data. \cite{tensors}

Just like vectors can alternatively characterized by the concept of a vector space, tensors can be abstractly defined as objects of an algebraic tensor product \cite{greub}. Given $K$ vector spaces $V_1, \dotsc, V_K$, their tensor product is a new and bigger vector space $V_1 \otimes \dotsb \otimes V_K$ together with a $K$-linear function $\otimes\colon V_1 \times \dotsb \times V_K \to V_1 \otimes \dotsb \otimes V_K$---called the tensor product of vectors---with the following universal property: any $K$-linear function $f\colon V_1, \dotsm, V_K \to W$ to any vector space $W$ may be decomposed, uniquely, as $f = \bar{f} \circ \otimes$, where $\bar{f}\colon V_1 \otimes \dotsb \otimes V_K \to W$ is a linear function that represents $f$ in tensor language. In other words, the tensor product makes it possible to represent $K$-linear functions, often unfamiliar, as linear functions, which are well understood. This can be summarized in the following commutative diagram.
\begin{equation}\label{eq:universal}
  \raisebox{-0.5\height}{\includegraphics{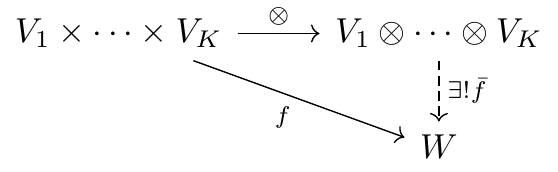}}
			\end{equation}

The tensor product of the vectors $v_1\in V_1, \dotsc, v_k\in V_K$ is often represented as $v_1 \otimes \dotsb \otimes v_k$, which is an order $K$ tensor. In computational terms, the tensor product can be implemented as a Kronecker product \cite{kron}. Any tensor that can be decomposed in a tensor product of vectors is called \emph{decomposable}. It is a consequence of the definition of the algebraic tensor product that any tensor $\mathcal{T}\in V_1 \otimes \dotsb \otimes V_K$ can be written as a sum of decomposable tensors, as in \eqref{eq:tensorrank}.
\begin{equation}\label{eq:tensorrank}
\mathcal{T} = \sum_{r=1}^{R} v_{1,r} \otimes \dotsb \otimes v_{K,r}.
\end{equation}

There are many ways of writing a tensor this way, but the well-ordering principle implies that, for any given tensor, there exists a minimum number $R$ for which this is possible. This is called the rank of the tensor and will be represented as $\rank \mathcal{T}$. The rank of the \emph{zero tensor} is defined as zero, being the only tensor with such rank. Decomposable tensors have rank either one or zero. As an abuse of language, the paper shall refer to decomposable tensors also as rank-one tensors. This will not create confusion and is quite usual in tensor literature. When $K = 2$, the notion of tensor rank reduces to that of matrix rank. 

Other property of interest is the dimension of the algebraic tensor product:
\begin{equation}
	\dim  V_1 \otimes \dotsb \otimes V_K = (\dim V_1) \dotsm (\dim V_K),
\end{equation}
that is, the dimension of the resulting vector space is the product of the dimensions of the individual vector spaces. This is consistent with the multi-index representation of tensors.

In classical tensor literature, lower and upper indexes in the coordinate representation are treated differently. This is to facilitate the introduction of operations similar to the product of matrices. Upper indexes are related to column vectors, in the way of $(v)^{i}$, while row vectors are represented by a lower index: $(w)_j$. A general tensor may be represented as $(\mathcal{T})^{i_1\dots i_K}_{j_1 \dots j_L}$. 

In this paper the following convention will be made: given tensors $\mathcal{T}_1$ and $\mathcal{T}_2$, indexed as $(\mathcal{T}_1)^{i_1 \dots i_K}_{j_1 \dots j_P}$ and $(\mathcal{T}_2)^{i_1 \dots i_P}_{j_1 \dots j_L}$, the product $\mathcal{T}_1 \mathcal{T}_2$ is defined by the coordinates
\begin{equation}\label{eq:convention}
	(\mathcal{T}_1 \mathcal{T}_2)^{i_1 \dots i_K}_{j_1 \dots j_L} = \sum_{k_1,\dotsc,k_P} (\mathcal{T}_1)^{i_1 \dots i_K}_{k_1 \dots k_P} (\mathcal{T}_2)^{k_1 \dots k_P}_{j_1 \dots j_L}.
\end{equation}

Notice that this is defined only for when the number of lower indexes of $\mathcal{T}_1$ equals the number of upper indexes of $\mathcal{T}_2$. In particular, this convention implies, for the column vector $(v)^{i}$ and the row vector $(w)_j$, that the products $w v$ and $v w$ are, respectively, the inner and outer products of $v$ and $w$, which is consistent with matrix notation.

Another important operation is the tensor product of tensors, which takes the two tensors $(\mathcal{T}_1)^{i_1 \dots i_K}_{j_1 \dots j_P}$ and $(\mathcal{T}_2)^{k_1 \dots k_Q}_{\ell_1 \dots \ell_L}$ and result in a higher order tensor $\mathcal{T}_1 \otimes \mathcal{T}_2$ given by
\begin{equation}\label{eq:convention}
	(\mathcal{T}_1 \otimes \mathcal{T}_2)^{i_1 \dots i_K k_1 \dots k_Q}_{j_1 \dots j_P \ell_1 \dots \ell_L} = (\mathcal{T}_1)^{i_1 \dots i_K}_{j_1 \dots j_P}(\mathcal{T}_2)^{k_1 \dots k_Q}_{\ell_1 \dots \ell_L},
\end{equation}
that it, given by juxtaposition of their coordinates via multiplication.

\subsection{Rank-one approximation}

It is possible to naturally extend norms and inner products of vectors to corresponding norms of tensors. The details of how to compute them are presented on Appendix \ref{sec:tensornorms}. These notions allow for the proposition of approximation problems. For example, it is possible to pose the rank $R$ approximation problem: given a tensor $\mathcal{T}$, find a tensor $\mathcal{X}$ that solves the problem
\begin{equation}\label{eq:estimation}
	\min_{\mathcal{X}} \|\mathcal{T} - \mathcal{X}\|, \quad \text{s. t.}\quad \rank{\mathcal{X}} \le R.
\end{equation}

This problem has the potential to decrease the number of parameters necessary to represent the tensor, but it is, in general, not well posed and may not have a solution \cite{illposed}. One case when this always has a solution is with order $2$ tensors---that is, matrices. But, for general order $K$ tensors, the case where it is always guaranteed to have a solution is when $R = 1$, the so called rank-one approximation. This fact follows from the next proposition.

\begin{thm}
The set $D = \{\mathcal{X}\colon \rank \mathcal{X} \le 1\}$ of decomposable tensors is closed.
\begin{proof}
Proof given by \cite{illposed} in Proposition 4.2.
\end{proof} 
\end{thm}

It is, therefore, advantageous to base signal processing algorithms over the problem of rank-one approximations, as these algorithms will most likely be well-behaved.

\section{The Volterra Series}
Traditionally, the Volterra series is treated as an universal approximator for nonlinear systems. It can be described as a polynomial representation of the system, closely related to a Taylor series representations. Explicitly, given a nonlinear system with input signal $u(i)$ and output $y(i)$, its Volterra series representation is the sum
\begin{equation}
	y(i) = y^0 + y^1(i) + y^2(i) + y^3(i) + \dotsb,
\end{equation}
where $y^0$ is a constant and each $y^k(i)$ is called the order $K$ homogeneous component of the system and is given by
\begin{equation}
	y^k(i) = \sum_{i_1, \dotsc, i_k} \mathcal{H}_k(i_1, \dotsc, i_k) u(i - i_1) \dotsm u(i - i_k),
\end{equation}
in which $\mathcal{H}_k(i_1, \dotsb, i_k)$ is a set of parameters called the order $k$ Volterra kernel.

The truncated Volterra series is obtained by limiting this series both in order and in time. This results in
\begin{equation}
	y(i) = y^0 + y^1(i) + \dotsb + y^K(i)
\end{equation}
and
\begin{equation}\label{eq:volt}
	y^k(i) = \sum_{i_1, \dotsc, i_k = 0}^M \mathcal{H}_k(i_1, \dotsc, i_k) u(i - i_1) \dotsm u(i - i_k),
\end{equation}
where $K$ is the order of the representation and $M$ its memory parameter. If follows from the Stone--Weierstrass theorem that the truncated Volterra series is an universal approximator---it can arbitrarily approximate any continuous nonlinearity, given that $K$ and $M$ are big enough.

Although this is an interesting property, the representational complexity, the number of the parameters necessary to describe an order $k$ Volterra kernel is $O(M^k)$, or, by exploring symmetry, $O\left(M - k + 1 \choose k\right)$. This high complexity gets transfered to Volterra-based algorithms, something that has historically barred their use in applications.

To deal with this problem, a common approach in the literature is to assume some restriction in the series and/or kernels. For example, for second order kernels, one possible approach is to truncate some of the diagonals of $\mathcal{H}_2(i,j)$ \cite{svolt,svolt2}, which allows the the computation to be written as 
\begin{equation}\label{eq:simpvolt}
y^2(i) = \sum_{d=0}^{D-1}\sum_{i_1}^{M-1} \mathcal{H}_2(i_1, i_1 + d) u(i - i_1) u(i - i_1 - d),
\end{equation}
where the $D$ is the number of diagonal utilized. When $D = 1$, this uses only the main diagonal, with $M$ parameters, resulting in a model called the Power Filter \cite{pfilter}.

Another approach is to start from tensor representations of the Volterra kernel. The multi-index structure of the Volterra kernel is quite obvious and it may as well have been represented as $(\mathcal{H}_k)^{i_1 \dots i_k}$, making evident that it is a tensor. In a more conceptual way, one may realize that the $k$-th order component may as well be computed as a $k$-linear transformation. First, define the input vector as the row vector $(u_i)_j = u(i - j + 1)$, or:
\begin{equation}
	u_i = 
	\begin{bmatrix}
		u(i) & u(i-1) & \dots & u(i - M + 1)
	\end{bmatrix}
\end{equation}

Then, notice how \eqref{eq:volt} may be rewritten, using the conventions of \eqref{eq:convention}, as 
\begin{equation}\label{eq:small}
y^k(i) = u_i^{\otimes k} \mathcal{H}_k,
\end{equation}
where $ u_i^{\otimes k}\triangleq u_i \otimes \dotsb \otimes u_i$ is the tensor power of $u_i$---the tensor product of $u_i$ with itself $k$ times. Eq. \eqref{eq:small} could be seen as a function $f\colon \mathbb{C}^M \times \dotsb \times \mathbb{C}^M \to \mathbb{C}$ given by 
\begin{equation} 
	f(v_1, \dotsc, v_k) = (v_1 \otimes \dotsb \otimes v_k) \mathcal{H}_k,
\end{equation}
which is, by the multilinearity of the tensor product, a $k$-linear function. So, from the universal property \eqref{eq:universal}, it is natural to expect the Volterra series to have a tensor representation. 

This representation has been used, for symmetric tensors, in \cite{favier}, to reduce the representational complexity of the series. 

In this paper, the restricting hypothesis will be simply the one of decomposability of kernel.

\subsection{Decomposable Model}

Consider the Volterra kernel tensor $\mathcal{H}_K$ of order $K$ and assume that it is a decomposable tensor, that it, that there exists vectors $w_1, \dots, w_K\in \mathbb{C}^M$ such that 
\begin{equation}
	\mathcal{H}_K = w_1 \otimes \dotsb \otimes w_K.
\end{equation}

A kernel like this will be said to satisfy the decomposable model---also called the Simple Multilinear Model (SML) \cite{eusipco2016}. This model allows for great gains in computational complexity, as it can be seen by carrying out the following computations:
\begin{align}
	y^K(i) &= u_i^{\otimes k} \mathcal{H}_k = (\underbrace{u_i \otimes \dotsb \otimes u_i}_{K\text{ times}})(w_1 \otimes \dotsb \otimes w_K)\nonumber\\
		&= \sum_{i_1,\dots,i_K=0}^M u(i - i_1) \dotsm u(i - i_K) w_1(i_1) \dotsm w_K(i_K)\nonumber\\
		& = \sum_{i_1 = 1}^M w_1(i_1) u(i - i_1) \dotsm \sum_{i_K = 1}^M w_1(i_1) u(i - i_1)\nonumber\\
		& = (u_i w_1) \dotsm (u_i w_K).
\end{align}

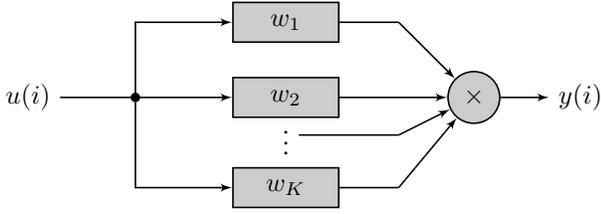
\begin{figure}[!t]
\pgfdeclarelayer{background layer}
\pgfdeclarelayer{foreground layer}
\pgfsetlayers{background layer,main,foreground layer}
\centering
\begin{tikzpicture}[auto, node distance=0.5cm,>=latex', semithick]

\node [input, label = left:$u(i)$] (input) {};
\node [circle, fill = black, draw = black, minimum size=3pt, inner sep=0pt, node distance = 1.0 cm, right of = input] (inputbranch) {};
\node [block, right of=input, node distance = 3.0 cm] (afilter) {$w_2$};
\node [sum, right of = afilter, node distance = 2.5 cm] (diff) {$\times$};

\node [block, above of=afilter, node distance = 1.0 cm] (system) {$w_1$};
\node [below of = afilter, node distance = 0.5 cm] (dot) {$\vdots$};
\node [block, below of=dot, node distance = 0.7 cm] (wk) {$w_K$};

\node [coordinate, right of = system, node distance = 1.5 cm] (o1) {$\vdots$};
\node [coordinate, right of = wk, node distance = 1.5 cm] (ok) {$\vdots$};
\node [coordinate, right of = dot, node distance = 1.5 cm] (od) {$\vdots$};

\node [output, right of=diff,node distance = 1.0 cm, label = right:$y(i)$] (out) {};

\draw [->] (input) -- node {} (afilter);
\draw [->] (afilter) -- node {} (diff);
\draw [->] (inputbranch) |- node {} (system);
\draw [->] (inputbranch) |- node {} (wk);
\draw [-] (system) -- node {} (o1);
\draw [->] (o1) -- node {} (diff);
\draw [-] (wk) -- node {} (ok);
\draw [->] (ok) -- node {} (diff);
\draw [-] (dot) -- node {} (od);
\draw [->] (od) -- node {} (diff);
\draw [->] (diff) -- node {} (out);
\end{tikzpicture}
\caption{Block diagram of the decomposable model.}
\label{fig:block}
\end{figure}

This means that the output may be computed as a product of the outputs of $K$ FIR linear systems. This is represented in Fig. \ref{fig:block}. In terms of computational complexity, this result in something based on $O(KM)$, an exponential reduction when compared to the original complexity of the Volterra series, which was of $O(M^K)$. This, together with the well-posedness of the estimation problem should make for a good basis onto which develop adaptive algorithms.

\section{Estimation Theory of the Decomposable Model}

Suppose one wants to estimate the best rank-one approximation of the homogeneous component $y_K(i)$ of the Volterra series by estimating the approximation of its kernel $\mathcal{H}_K$. The problem can be posed in the following way: one is given a random signal $\mathbf{d}$ (the desired signal) and a random $1\times M$ vector $\mathbf{u}$ (the regressor). One want to find the $K$ order decomposable Volterra kernel $\mathcal{W}$ that best estimates $\mathbf{d}$, in the mean-square sense. This can be put in terms of the constrained optimization problem below.

\begin{equation}\label{eq:estimation}
	\min_{\mathcal{W}} \mathbb{E}\left|\mathbf{d} - \mathbf{u}^{\otimes K} \mathcal{W}\right|^2, \quad \text{s. t.}\quad \rank \mathcal{W} \le 1
\end{equation}

This problem has a cost function given by $J(\mathcal{W}) \triangleq \mathbb{E}\left|\mathbf{d} - \mathbf{u}^{\otimes K} \mathcal{W}\right|^2$. This cost function is called the Mean Square Error (MSE), while the signal $\mathbf{e}\triangleq \mathbf{d} - \mathbf{u}^{\otimes K} \mathcal{W}$. The decomposability constraint can be included in this function by using $\mathcal{W} = w_1 \otimes \dotsb \otimes w_K$, for vectors $w_1, \dotsc, w_K$ ($M\times 1$). It also will be convenient to also define the vector $w$ ($KM \times 1$) built by vertically stacking the vectors $w_1, \dotsc, w_K$. 

It is possible to explicitly develop the MSE as\footnote{The star operation ($^*$) on a tensor $(\mathcal{T})^{i_1 \dots i_K}_{j_1 \dots j_L}$ consists of changing lower indexes for upper indexes and taking the complex conjugate of the entries: $(\mathcal{T}^*)^{i_1 \dots i_L}_{j_1 \dots j_K} = \overline{(\mathcal{T})^{j_1 \dots j_K}_{i_1 \dots i_L}}$.}
\begin{align}\label{eq:mse_eval}
J(w) &= \mathbb{E}|\mathbf{e}|^2 = \mathbb{E}\left[[\mathbf{d} - \mathbf{u}^{\otimes K}\mathcal{W}]^* [\mathbf{d} - \mathbf{u}^{\otimes K}\mathcal{W}]\right] \nonumber\\
&=  \mathbb{E}|\mathbf{d}|^2 - \mathcal{W}^*\mathbb{E}[\mathbf{d} \mathbf{u}^{\otimes K*}] - \mathbb{E}[\mathbf{d}^* \mathbf{u}^{\otimes K}] \mathcal{W} \nonumber\\
&\quad + \mathcal{W}^* \mathbb{E}[\mathbf{u}^{\otimes K*} \mathbf{u}^{\otimes K}] \mathcal{W}.\nonumber\\
& =  R_d - \mathcal{W}^* \mathcal{R}_{u^K d}^* - \mathcal{R}_{u^K d} \mathcal{W} + \mathcal{W}^* \mathcal{R}_{u^K} \mathcal{W},
\end{align}
where the correlation parameters were defined as
\begin{gather}
\mathcal{R}_{u^K} = \mathbb{E}[\mathbf{u}^{\otimes K*} \mathbf{u}^{\otimes K}],\label{eq:ruk}\\ 
\mathcal{R}_{u^K d} = \mathbb{E}[\mathbf{u}^{\otimes K} \mathbf{d}^*]  = \mathcal{R}_{du^K}^*,\\
R_d = \mathbb{E}|\mathbf{d}|^2.
\end{gather}

The minimum of this function is achieved at a point where the gradient, over $w$, is zero. This gradient has a block structure, each depending only on one $w_s$:
\begin{equation}
\nabla J(w) = \begin{bmatrix}
\frac{\partial J}{\partial w_1} & \frac{\partial J}{\partial w_2} & \cdots & \frac{\partial J}{\partial w_K}
\end{bmatrix}.
\end{equation}

Each block may be computed as in the next proposition. The derivatives used are Wirtinger derivatives.

\begin{prop}
The gradient of the MSE function with respect to $w_s$ can be computed as
\begin{align}\label{eq:grad}
\frac{\partial J}{\partial w_s} = [-\mathcal{R}_{u^K d} + \mathcal{W}^* \mathcal{R}_{u^K}] \mathcal{W}^{(s)},\\
\text{with } \mathcal{W}^{(s)} = (w_1 \otimes \dotsb \otimes \widehat{w_s }\otimes \dotsb \otimes w_K),\label{eq:ident}
\end{align}
where $\widehat{w_s}$ implies that $w_s$ has been substituted for the identity matrix $I_M$ of order $M$ in the product. 
\begin{proof}
The following tensor indexations shall be used:
\begin{gather}
(\mathcal{R}_{u^ K})^ {i_1,\dotsc,i_K}_{j_1,\dotsc,j_K}=\mathbb{E}\left[(\mathbf{u}^{\otimes K *})^ {i_1,\dotsc,i_K}(\mathbf{u}^{\otimes K})_{j_1,\dotsc,j_K}\right]\\
\quad(\mathcal{R}_{u^Kd})_{j_1,\dotsc,j_K} = \mathbb{E}\left[(\mathbf{u}^{\otimes K })_{j_1,\dotsc,j_K} \mathbf{d}\right].
\end{gather}

Through this, we can write
\begin{equation}
\mathcal{R}_{u^Kd}\mathcal{W} = \sum_{j_1,\dotsc,j_K}(\mathcal{R}_{u^Kd})_{j_1,\dotsc,j_K}\prod_{\ell = 1}^ K (w_\ell)^{j_\ell},
\end{equation}
where $(w_\ell)^{j_\ell}$ is the $j_\ell$-th coordinate of $w_\ell$, and
\begin{equation}
\mathcal{W}^*\mathcal{R}_{u^K}\mathcal{W} = \sum_{\substack{i_1,\dotsc,i_K\\j_1,\dotsc,j_K}}\prod_{p=1}^K (w_p)^{i_{p}*}(\mathcal{R}_{u^K})^{i_1,\dotsc,i_K}_{j_1,\dotsc,j_K}\prod_{\ell = 1}^K (w_\ell)^{j_{\ell}}.
\end{equation}

The other terms from \eqref{eq:mse_eval} involve only the conjugates of the entries of $w$, so their Wirtinger derivatives become zero \cite{bk:sayed}.

The gradient over the vector $w_{s}$ is given by the derivatives over each of its components, which, by introducing the symbol $\delta^i_j$---called the Kronecker delta---that evaluates to $0$ if $i \ne j$ and to $1$ if $i = j$, results in:
\begin{equation}\label{eq:indexes}
\frac{\partial (\mathcal{R}_{u^Kd}\mathcal{W})}{\partial(w_s)^{j_q}} = \sum_{j_1,\dotsc,j_K}(\mathcal{R}_{u^Kd})_{j_1,\dotsc,j_K}\prod_{\ell\ne s} (w_\ell)^{j_\ell}\delta^{j_s}_{j_q}.
\end{equation}

As $j_q$ take values from $1$ through $M$, \eqref{eq:indexes} indeed renders the vector \eqref{eq:grad}. This can be shown as follows: the Kronecker delta $\delta^i_j$ indexes the identity matrix, that is, $(I_M)^i_j = \delta^i_j$. In \eqref{eq:indexes}, the delta works as if occupying the positions of the coordinates $(w_s)^{j_q}$---that is, the coordinates of an identity $I_M$ are occupying the positions of the coordinates of the vector $w_s$. This is valid for any index $j_q$, thus, when reconstructing the tensor form implied by the coordinates computed in \eqref{eq:indexes}, the substitution with $I_M$ in \eqref{eq:ident} must be made:
\begin{equation}
\frac{\partial (\mathcal{R}_{u^Kd}\mathcal{W})}{\partial w_s} =  \mathcal{R}_{u^Kd} (w_1 \otimes \dotsb \otimes\widehat{ w_s}\otimes \dotsb \otimes w_K) = \mathcal{R}_{u^Kd} \mathcal{W}^{(s)}.
\end{equation}

For $\mathcal{W}^*\mathcal{R}_{u^K}\mathcal{W}$, we have, remembering we do not derivate the conjugates,
\begin{equation}\label{eq:wrw}
\frac{\partial (w^*\mathcal{R}_{u^K}w)}{\partial(w_s)^{j_q}} = \sum_{\substack{i_1,\dotsc,i_K\\j_1,\dotsc,j_K}}\prod_{p=1}^K (w_p^*)_{i_{p}}(\mathcal{R}_{u^K})^{i_1,\dotsc,i_K}_{j_1,\dotsc,j_K}\prod_{\ell\ne s} (w_\ell)^{j_{\ell}}\delta^{j_s}_{j_q}.
\end{equation}

Under the same argument, 
\begin{equation}
\frac{\partial (\mathcal{W}\mathcal{R}_{u^K}\mathcal{W}^*)}{\partial w_s} = \mathcal{W}^* \mathcal{R}_{u^K} \mathcal{W}^{(s)}.
\end{equation}

Therefore, one combines those two terms to get
\begin{align}
\frac{\partial J}{\partial w_s} &= \frac{\partial}{\partial w_s} (R_d - \mathcal{W}^* \mathcal{R}_{u^K d}^*  - \mathcal{R}_{u^K d} \mathcal{W} + \mathcal{W}^* \mathcal{R}_{u^K} \mathcal{W})\nonumber\\
&= -\frac{\partial (\mathcal{R}_{u^Kd}\mathcal{W})}{\partial w_s}  + \frac{\partial (\mathcal{W}\mathcal{R}_{u^K}\mathcal{W}^*)}{\partial w_s} \nonumber\\
&= [-\mathcal{R}_{u^K d} + \mathcal{W}^* \mathcal{R}_{u^K}]\mathcal{W}^{(s)}.
\end{align}
\end{proof}
\end{prop}

In a critical point $\nabla J(w) = 0$ of this surface, $w$ satisfies a series of polynomial equations in its entries reminiscent of the normal equations of linear estimation theory:
\begin{equation}\label{eq:normal}
[-\mathcal{R}_{u^K d} + \mathcal{W}^* \mathcal{R}_{u^K}] \mathcal{W}^{(s)} = 0, \quad 1\le s \le K.
\end{equation}

One can verify the \emph{possibility}---because this surface is multi-modal---of minimality of a certain $w$ by testing it in this equation, but a solution via direct methods may be difficult. It should be more viable to use iterative algorithms.

\section{The Steepest Descent Algorithm}

The steepest descent algorithm can be used to find the minimum of \eqref{eq:mse_eval}. This algorithm is based on the recursion
\begin{equation}
	w_i = w_{i-1} -\mu [\nabla J(w_{i-1})]^*.
\end{equation}

In terms of the individual vectors, this may be rewritten as
\begin{equation}\label{eq:steep}
	w_{s,i} = w_{s,i-1} + \mu \mathcal{W}^{(s)*}[\mathcal{R}_{du^K} - \mathcal{R}_{u^K}\mathcal{W}].
\end{equation}

Due to the nonlinear nature of the algorithm, there are some issues with the initializations of the $w_s$ parameters. First, the should not be initialized all at zero, as \eqref{eq:steep} shows that this would make the parameters stay at zero through all of the iteration process. Another bad initialization is to set all of them at the same initial values. The recursion shows that this would lead to them being adapted exactly in the same way, never being able to diverge from one another. Initializing the parameters as pairs of opposite vectors leads to a similar problem with alternating iterations. Aside from random initializations, the previous discussion makes the heuristic argument to initialize the vectors in different scales. The following set of initializations has shown good results in our experiments.
\begin{gather}\label{eq:initial}
w_{s,-1} = \begin{bmatrix}2^{-s+1} & 0 & \dots & 0\end{bmatrix}, \text{ for } 1\le s \le K-1\\
w_{K,-1} = \begin{bmatrix}0 & 0 & \dots & 0\end{bmatrix}.
\end{gather}

\begin{figure}[!t]
\centering
\subfloat[$K =2$.]{
\includegraphics[clip = true, trim = 200 315 200 325, width=1.58in]{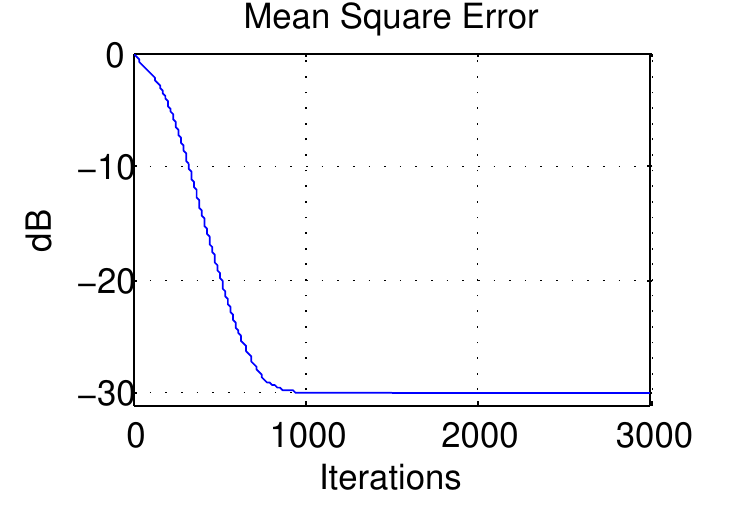}%
\label{fig:steep1}
}\hspace{2mm}
\subfloat[$K =3$.]{
\includegraphics[clip = true, trim = 200 315 200 325, width=1.58in]{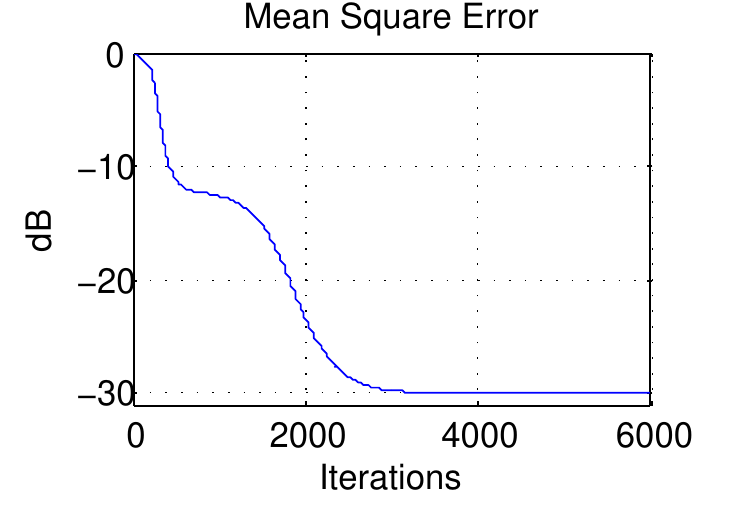}%
\label{fig:steep2}
}
\caption{MSE curves for the steepest descent algorithm.}
\label{fig:steep}
\end{figure}

This recursion was used to find a solution to the estimation problem with 
\begin{equation}
\mathbf{d} = \mathbf{u}^{\otimes K} \mathcal{H} + \mathbf{v},
\end{equation}
where $\mathbf{u}$ is a $1\times 10$ independent Gaussian vector, $\mathbf{v}$ is a signal independent from $\mathbf{u}$ with variance $\sigma^2 = 10^{-3}$ and the plant $\mathcal{H}$ is decomposable.. The resulting MSE curve is on Figs. \ref{fig:steep1} ($K = 2$) and \ref{fig:steep2} ($K=3$). It is possible to notice how the algorithm is able to reach the theoretical minimum MSE of $\sigma^2$. Moreover, it was possible to verify that the solutions found satisfy the normal equations \eqref{eq:normal}.

Although some good behavior of this algorithm can be attested by these simulations, in practice it is very hard for one to obtain the $\mathcal{R}$ parameters. So a recursion like \ref{eq:steep} is rarely used---the interest in it is a rather theoretical one. A solution to this problem is to compute real-time approximations to \eqref{eq:ruk}. This is called an adaptive solution. Not only is solves the problem of unavailability of the correlation parameters, it also gives the algorithm some tracking capabilities. When doing this to the linear estimation problem, the algorithm obtained is the classical LMS \cite{bk:sayed}. When done to the decomposable model, it should, therefore, lead to similar algorithms.

\section{Adaptive Algorithms}

As it is impractical to use the steepest descent algorithm, one finds an adaptive solution that implements a recursion that approximates \eqref{eq:steep}. This is done by computing a real-time approximation to the parameters \eqref{eq:ruk}.

\subsection{Rectangular Window Approximation}
A possible approximation is to use a rectangular window to estimate the correlation parameters---that is, the last $L$ samples of the signals are used in a average to compute the approximations
\begin{align}
&\widetilde{\mathcal{R}}_{u^K} = \frac{1}{L}\sum_{j=i-L+1}^i u_j^{\otimes K*} u_j^{\otimes K},\label{eq:ruka} \\
&\widetilde{\mathcal{R}}_{u^K d} = \frac{1}{L}\sum_{j=i-L+1}^i u_j^{\otimes K} d(j)^*  = \widetilde{\mathcal{R}}_{du^K}^*.\label{eq:rduka}
\end{align}

This allows for the computation of an estimate of the block-gradient \eqref{eq:grad}:
\begin{align}
\frac{\widetilde{\partial J}}{\partial w_s} &= [-\widetilde{\mathcal{R}}_{u^K d} + \mathcal{W}^* \widetilde{\mathcal{R}}_{u^K}] \mathcal{W}^{(s)}\nonumber\\
&= \frac{1}{L}\sum_{j=i-L+1}^i\left[-d(j)^* + \mathcal{W}^* u_j^{\otimes K*}\right]u_j^{\otimes K}\mathcal{W}^{(s)}.\label{eq:gradsum}
\end{align}

Equation \eqref{eq:gradsum} may be conveniently rewritten as follows. First define $y_s(j) \triangleq (u_j w_1) \widehat{\dotsb (u_j w_s) \dotsb} (u_j w_K)$, where the hat implies a factor being omitted, which implies $u_j^{\otimes K}\mathcal{W}^{(s)} = y_s(j) u_j$, then define three variables:
\begin{gather}
d_i \triangleq \begin{pmatrix}
d(i)\\
\vdots\\
d(i - L + 1)
\end{pmatrix}\, {\scriptstyle(L\times 1)},
\,\,
U_i^K \triangleq \begin{pmatrix}
u_i^{\otimes K}\\
\vdots\\
u_{i - L + 1}^{\otimes K}
\end{pmatrix}\, {\scriptstyle(L \times KM)},\nonumber\\
y_j \triangleq \begin{pmatrix}
y_1(j)&
\cdots&
y_K(j)
\end{pmatrix}\, {\scriptstyle(1\times K)}.
\end{gather}

Eq. \eqref{eq:gradsum} represents a block-gradient. For each $j$, the terms of this sum will be multiplied only by $y_s(j)$---the $s$-th element of $y_j$---which shows that the total gradient has a Kronecker structure. This follows from the fact that $u_j^{\otimes K}\mathcal{W}^{(s)} = y_s^w(j) u_j$. Therefore, write
\begin{equation}
\widetilde{\nabla}J = \frac{1}{L}\sum_{j=i-L+1}^i\left[-d(j)^* + \mathcal{W}^* u_j^{\otimes K*}\right] \left(y_j\otimes u_j\right).
\end{equation}

By using the variable
\begin{equation}\label{eq:ti}
T_i \triangleq \begin{pmatrix}
y_i\otimes u_i\\
\vdots\\
y_{i-L+1}\otimes u_{i-L+1}
\end{pmatrix}\, {\scriptstyle(L\times KM)},
\end{equation}
a compact expression for the instantaneous gradient is obtained:
\begin{align}\label{eq:estigrad}
\widetilde{\nabla}J = -\frac{1}{L}\left[d_i -  U_i^K \mathcal{W}\right]^*T_i =  -\frac{1}{L} e_i^* T_i,
\end{align}
where $e_i \triangleq d_i -  U_i^K \mathcal{W}$ ($L\times 1$). Finally, defining $y(j) \triangleq u^{\otimes K} \mathcal{W}$ and
\begin{equation}
y_i \triangleq \begin{pmatrix} y(i) & \cdots & y(i-L+1)\end{pmatrix}^T \quad (L\times 1),
\end{equation}
results in an expression for $e_i$ as
\begin{equation}
e_i = d_i - y_i.
\end{equation}

\subsection{LMS Algorithm}
When $L = 1$, Eq. \eqref{eq:estigrad} becomes
\begin{equation}\label{eq:lmsgrad}
\widetilde{\nabla}J = - e(i)^* (y_i \otimes u_i),
\end{equation}
where $e(i) \triangleq d(i) - u_i^{\otimes} \mathcal{W}$. When separated for each $w_s$, the LMS equation becomes
\begin{align}\label{eq:lms}
w_{s,i} = w_{s,i-1}+ \mu e(i) y_{s}(i)^*  u_i^*.
\end{align}

This is a Least Mean Squares (LMS) recursion.

\subsection{True-LMS Algorithm}
When using a general value of $L$, the recursion of the TRUE-LMS (also sometimes called the Dat  Reuse LMS) is derived:
\begin{align}
w_{i} = w_{i-1}+ \frac{\mu}{L} T_i^* e_i.
\end{align}

This can, by separating each $w_s$, be further simplified to
\begin{align}
w_{s,i} &= w_{s,i-1}+ \frac{\mu}{L}\sum_{j=i-L+1}^i [d(j) - u_j^{\otimes K}\mathcal{W}_{i-1}] y_s(j)^* u_j^*\nonumber\\
	&=  w_{s,i-1}+ \frac{\mu}{L} (y_{s,i} \circ U_i)^* e_i,
\end{align}
where
\begin{equation}
y_{s,i} \triangleq \begin{pmatrix}
y_s(i)\\
\vdots\\
y_s(i-L+1)
\end{pmatrix}\, {\scriptstyle(L\times K)},
\quad U_i \triangleq \begin{pmatrix}
u_i\\
\vdots\\
u_{i-L+1}
\end{pmatrix}\, {\scriptstyle(L\times M)}
\end{equation}
and $\circ$ denotes row-wise scalar multiplication.

\subsection{Stabilization of the Algorithms}
\label{ssec:stable}
\begin{figure}[!t]
\centering
\subfloat[The arrows point to signs of instability.The Figure is an ensemble average of $1.000$ realizations]{
\includegraphics[clip = true, trim = 115 315 130 315, width=8.5cm]{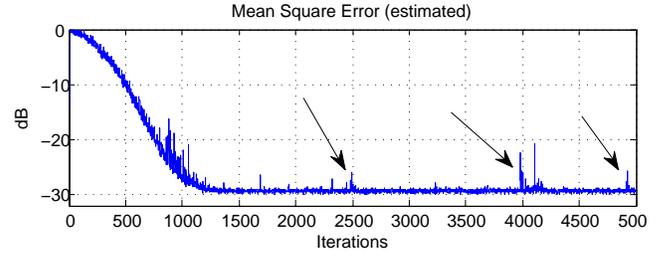}%
\label{fig:jumps}
}

\subfloat[A realization that becomes untable, using the same parameters from before.]{
\includegraphics[clip = true, trim = 115 315 130 315, width=8.5cm]{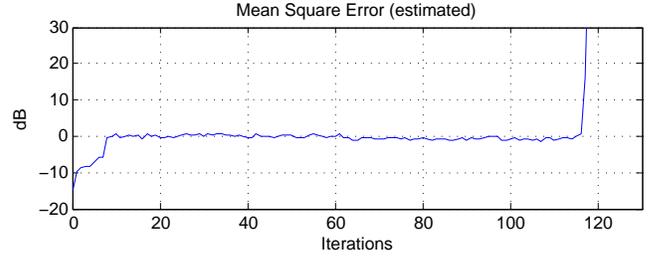}%
\label{fig:infinity}
}
\caption{Instability in the LMS algorithm.}
\label{fig:unstable}
\end{figure}

It has been known that nonlinearities in the update equations may lead to a nonzero probability of divergence for the algorithm, no matter how small is the step-size \cite{lmf}, whenever the probability distribution of the signals has infinite support. This is related to the fact that the algorithms are not globally asymptotically stable. An example of this phenomenon is illustrated in Fig. \ref{fig:unstable}. The big ``jumps'' in the MSE curve \ref{fig:jumps} shows the algorithm entering momentarily an area of instability. If the parameters got far enough from the stability region, the algorithm would diverge to infinity, as is shown on Fig. \ref{fig:infinity}.

For effective use of these algorithms, some normalization may be necessary. In the case of the LMS, a normalization may be performed by conditionally changing the update equation as in \eqref{eq:update2}:
\begin{align}\label{eq:update2}
w_{s,i} = \begin{cases}
w_{s,i-1}+\mu e(i) y_{s}(i)^*  u_i^* & \text{if } |y_{s}(i)| \le \text{MAX}\\
w_{s,i-1}+\mu e(i) u_i^* & \text{if } |y_{s}(i)| > \text{MAX}
\end{cases}.
\end{align}

\begin{figure}[!t]
\centering
\includegraphics[clip = true, trim = 115 315 130 315, width=8.5cm]{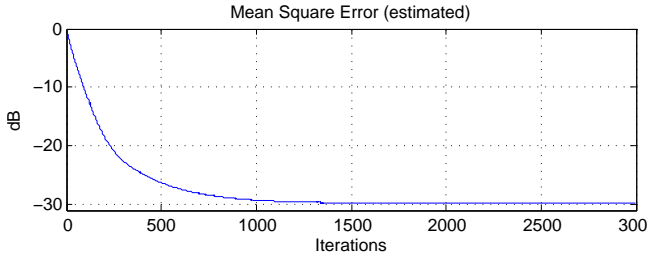}%
\caption{Simulation of recursion \eqref{eq:update2}, with an ensemble average of 1 million realizations.}
\label{fig:stable}
\end{figure}

This can be justified in a heuristic way. Whenever the algorithm starts diverging, $|y_{s}(i)|$ starts to increase. If it gets above a certain threshold, the update equation turns into the second line of \eqref{eq:update2}. If the step-size is small enough, this equation is approximately the classical LMS, an algorithm that is known to be stable in various situations. This stops the divergent behavior of the algorithm. At this point, the $w$ is not converging to the optimal value. This works as a gross convergence mode, that is guaranteed to be stable enough to allow $|y_{s}(i)|$ to decrease, entering again the the mode of fine convergence. Figure \ref{fig:stable} shows a simulation of this modified algorithm, averaged through 1 million realizations, with no signs of instability.

Additionally, a TRUE-LMS version of this modification can be done as 
\begin{align}\label{eq:trueupdate2}
w_{s,i} = \begin{cases}
w_{s,i-1}+ \frac{\mu}{L} (y_{s,i} \circ U_i)^* e_i & \text{if } |y_{s}(i)| \le \text{MAX}\\
 w_{s,i-1}+ \frac{\mu}{L} U_i^* e_i & \text{if } |y_{s}(i)| > \text{MAX}
\end{cases},
\end{align}
which works for the same reasons as before.

\subsection{Step Bounds}
For an initial description of the algorithms to complete, there must exist a method to choose the parameters of the algorithms. In this section a heuristic bound is obtained, based on comparisons with the classical linear LMS.

When computing the gradient for the classical algorithm, one obtains $\widetilde{\nabla}J = -e(i)^* u_i$. This leads to an instantaneous bound for the algorithm of $0 < \mu < 2/\|u_i\|^2$. The gradient obtained for the new algorithm is in \eqref{eq:lmsgrad}. Given that the bound depends only on this gradient, one would expect that an equivalent inequality for the new algorithm would be
\begin{equation}
0 < \mu < \frac{2}{\|y_i \otimes u_i\|^2} = \frac{2}{\|y_i\|^2 \|u_i\|^2}.
\end{equation}

This result is similar to the one obtained in \cite{rupp2}. Although important, instantaneous bounds may not be sufficient to choose the parameters of the algorithm. The classical LMS has a bound given by $0 < \mu < 2/(3 \tr R_u)$ that guarantees convergence in the MSE \cite{lmsstability}. Given that $R_u = \mathbb{E} \left[u_i^* u_i\right]$, an argument by comparison shows that the new nonlinear algorithm would depend on the trace of
\begin{equation}
\mathbb{E} \left[(y_i \otimes u_i)^* (y_i \otimes u_i)\right] = \mathbb{E}\left[ (y_i^* y_i) \otimes (u_i^* u_i)\right]
\end{equation}

In fact, simulations show that, for the exact gradient algorithm, this is exactly the parameter that governs convergence. If $\lambda_\text{max}$ is the biggest eigenvalue of $\lim_{i\to\infty}\mathbb{E}\left[ (y_i^* y_i) \otimes (u_i^* u_i)\right]$, then the steepest descent algorithm on \eqref{eq:steep} with initial conditions \eqref{eq:initial} converges to its global minimum whenever $0 < \mu < 2/\lambda_\text{max}$. This is analogous to the linear steepest descent. \cite{bk:sayed}

For the adaptive algorithms, since $y_i$ is not a stationary signal, this is still not an universal bound. To make it so, one should consider the signal after convergence, since, at that point, the weight vectors $w_1, \dotsc, w_K$ should be stationary, which makes $y_i$ an stationary signal, given that $u_i$ is also one. Under these conditions, the bound should be
\begin{equation}\label{eq:bound}
0 < \mu < \lim_{i\to \infty}\frac{2}{\alpha \tr \mathbb{E}\left[ (y_i^* y_i) \otimes (u_i^* u_i)\right]},
\end{equation}
where $\alpha$ is a constant. Due to the complexity involved in determining analytically the value of $\alpha$, which must take into account the modified algorithms \eqref{eq:update2} and \eqref{eq:trueupdate2}, an experimental approach was used, which pointed to a possible value of $\alpha = 3^K$, which depends on the order of nonlinearity $K$. This dependence can be explained by noting that higher $K$ yields a noisier convergence, which may put the algorithms in a unstable region, if the step-size is too big. One should note that in the linear case ($K = 1$) this bound reduces to the classical LMS bound.

It is also possible to give an estimate to the MAX parameter:
\begin{equation}\label{eq:max}
\text{MAX} = \lim_{i\to \infty} (K + 1) \sqrt{\mathbb{E}\left|u_i^{\otimes K} \mathcal{W}_i \right|^2}.
\end{equation}

This value was also obtained mostly through experimental verification, but heuristic reasoning is also possible. The $\sqrt{\mathbb{E}\left|u_i^{\otimes K} \mathcal{W}_i \right|^2}$ factor appears because the parameter MAX gives a condition on $|y_s(i)|$, so it should depend on the power of the output $y(i)$. The factor $K+1$ appears due to the experimental observation that the bigger the $K$, the noisier was the convergence, therefore a bigger value of MAX is necessary to allow this natural convergence behavior to be expressed, otherwise the algorithm would not converge to the lowest possible minimum.

These results should be valid for both the LMS and the TRUE-LMS, while, as it will be shown in simulations, they may be conservative choices for the latter, as it has better stability properties.In fact, It can be verified experimentally that for reasonably sized $L$ the ideal value of MAX for the TRUE-LMS is roughly half of \eqref{eq:max}.

\subsection{Computational Complexity}

Table \ref{tb:truelms} shows an efficient implementation of the TRUE-LMS algorithm, together with the necessary number of operations in each step of the process. It uses the following auxiliary variables:

\begin{gather}
y_{o\ell s}(i) \triangleq (u_{i-\ell+1} w_s)\\
y_{\ell s}(i) \triangleq y_{o\ell 1}(i)\dotsm \widehat{y_{o\ell s}(i)} \dotsm y_{o\ell K}(i)\\
y_{\ell}(i) \triangleq y_{\ell K}(i) y_{o\ell K}(i).
\end{gather}

\begin{table}[t]
  \centering
\caption{Implementation of the TRUE-LMS.}
  \label{tb:truelms}
  \begin{tabular}{c|cc}
  \hline
   & $+$ or $-$ & $\times$\\
  \hline
  $\forall s, \ell, y_{o\ell s}(i)$ & $LK(M-1)$  & $LKM$\\
  \hline
  $\forall s, \ell, y_{\ell s}(i)$ & $0$  & $LK(K-2)$\\
  \hline
  $\forall \ell, y_{\ell}(i)$ & $0$  & $L$\\
  \hline
  $e_i$ & $L$ & $0$\\
  \hline
  $T_i$ & $0$ & $LKM$\\
  \hline
  $Ti^{^*} e_i$ & $(L-1)KM$ & $LKM$\\
  \hline
  $w_{s}+ \frac{\mu}{L} Ti^{^*} e_i$ & $KM$ & $KM$\\
  \hline
  Total & \tiny$2 L K M - LK + L$ & \tiny$3 L K M + L K^2 + K M - 2 L K + L$\\
  \hline
  \end{tabular}
\end{table}

In Table \ref{tb:lms} and implementation of the LMS is shown. Some optimizations in relation to the TRUE-LMS are possible, which makes it more efficient than one would expect by using $L = 1$ in the previous table.

\begin{table}[t]
  \centering
\caption{Implementation of the LMS.}
  \label{tb:lms}
  \begin{tabular}{c|cc}
  \hline
   & $+$ or $-$ & $\times$\\
  \hline
  $\forall s, y_{o s}(i)$ & $K(M-1)$  & $KM$\\
  \hline
  $\forall s, y_{s}(i)$ & $0$  & $K(K-2)$\\
  \hline
  $y(i)$ & $0$  & $1$\\
  \hline
  $e(i)$ & $1$ & $0$\\
  \hline
  $\left[\mu e(i)\right] u_i^*$ & $0$ & $M+1$\\
  \hline
  $\forall s,  \left[\mu e(i) u_i^*\right] y_s(i)'$ & $0$ & $K$\\
  \hline
  $\forall s, w_{s}+ \mu e(i) y_{s}(i)^*  u_i^*$ & $KM$ & $0$\\
  \hline
  Total & \tiny$2 K M - K + 1$ & \tiny$K M + K^2 + M - K + 2$\\
  \hline
  \end{tabular}
\end{table}

\section{Simulations}

Some experiments with the algorithm were run. In all of these, the input vector $u_i$ was chosen to have a delay-line structure as in 
\begin{equation}
u_i = 
\begin{bmatrix}
u(i) & u(i - 1) & \dots & u(i - M + 1)
\end{bmatrix},
\end{equation}
with $u(i)$ an i.i.d. signal sampled from the normal distribution $\mathcal{N}(0, 1)$. 

The desired signal follows a ``system identification'' model, with
\begin{equation}
d(i) = u_i^{\otimes K} \mathcal{W}_o + v(i),
\end{equation}
where $\mathcal{W}_o$ is the plant to be identified and $v(i)$ is an i.i.d. signal sampled from the distribution $\mathcal{N}(0, \sigma_v^2)$ independent from $u(i)$.

All the adaptive algorithms simulated here are the stabilized versions \ref{eq:update2} and \ref{eq:trueupdate2}. They will be referred hereafter as SML.

\subsection{Decomposable plants}
\label{ssec:bestcase}
The simulations in this section use a decomposable plant, which represents the best case scenario for the SML algorithms. They were put to run against the Volterra-LMS and Wiener-LMS algorithms \cite{bk:ogunfunmi}. The Volterra algorithm is based on \eqref{eq:volt} and the Wiener one is a modified version in which the resulting regressor is statistically orthonormal. They both are linear-in-the-parameters algorithms, in contrast with the SML ones, which are not.

Simulations were done for $K = 2$ and $K = 3$ and they were repeated through $1.000$ and $10.000$ realizations, respectively. The plants were chosen randomly and normalized to result in unitary power. The memory parameter was chosen as $M = 10$ for all cases.

\begin{figure}[!t]
\centering
\subfloat[$K=2$ and $\sigma_v^2 = 10^{-3}$.]{
\includegraphics[clip = true, trim = 200 315 200 315, width=1.58in]{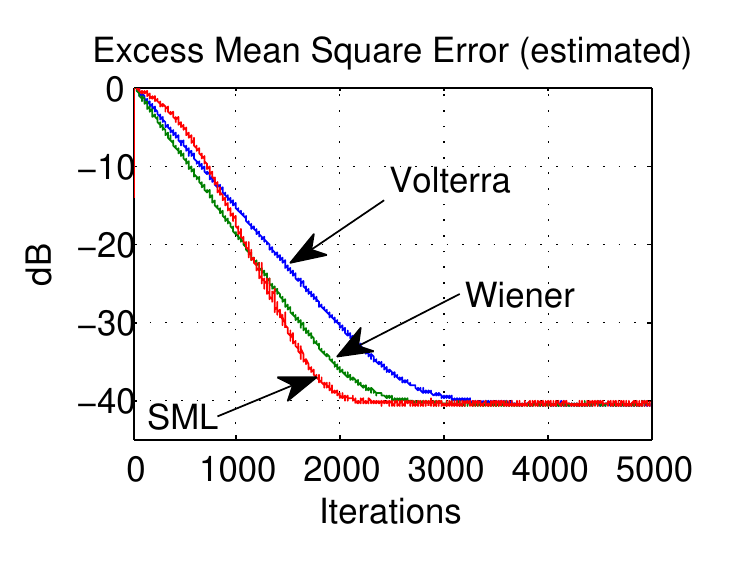}%
\label{ploti}
}\hspace{2mm}
\subfloat[$K=2$ and $\sigma_v^2 = 10^{-6}$, but with a different plant.]{
\includegraphics[clip = true, trim = 200 315 200 315, width=1.58in]{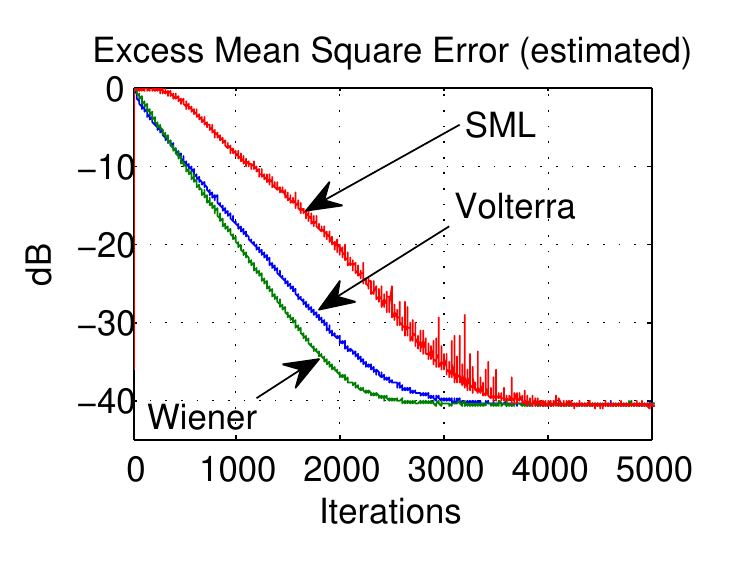}%
\label{plotii}
}

\subfloat[$K=3$ and $\sigma_v^2 = 10^{-6}$.]{
\includegraphics[clip = true, trim = 115 315 130 315, width=8.5cm]{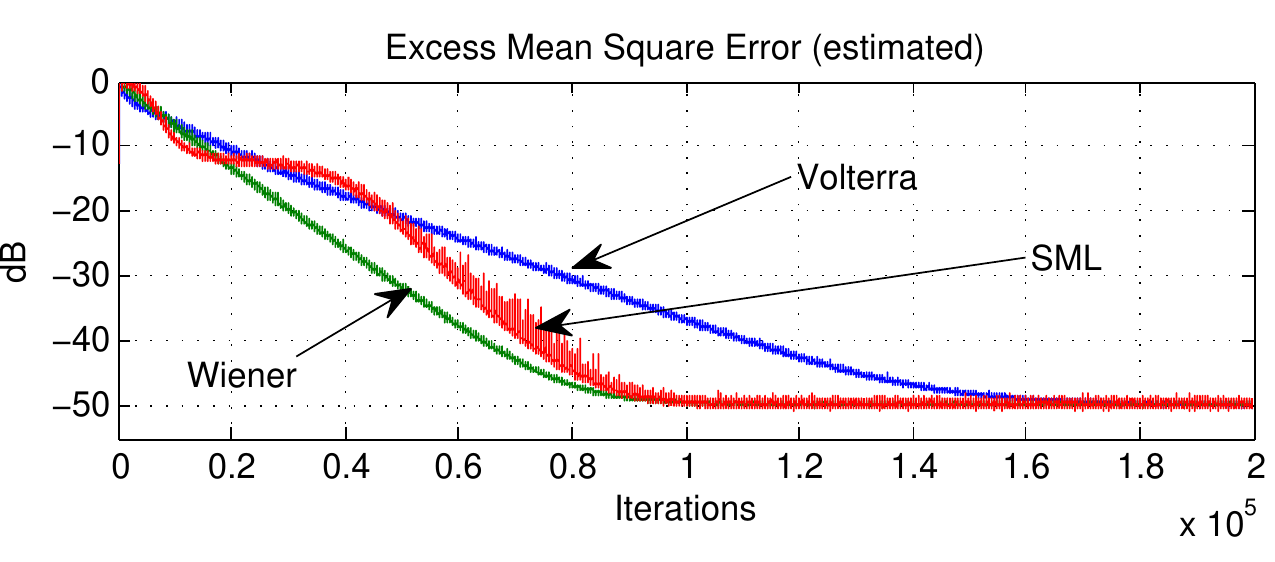}%
\label{plotiii}
}
\caption{Curves of the estimated EMSE showing the performance of the algorithms while identifying decomposable plants.}
\label{fig:plot}
\end{figure}

The algorithms were first simulated on an order $K=2$ plant with a noise of $\sigma_v^2 = 10^{-3}$  and only the SML-LMS algorithm was tested. The Excess Mean Square Error\footnote{Formally defined as $\mathbb{E}|(\mathcal{W}_o - \mathcal{W}_i)\mathbf{u}|^2$.} (EMSE) curves are on Fig. \ref{ploti}. The algorithms all have similar performances, but a difference starts to show after a few iterations, when the SML starts to converge faster and shows itself to be the first to reach convergence.

Fig. \ref{plotii} shows a similar scenario, with $\sigma_v^2 = 10^{-6}$; the only thing that was changed was the plant $w_o$. It is still decomposable and normalized plant, just a different one. This figure shows how the SML performance may be dependent on the plant being identified.

Fig. \ref{plotiii} shows a simulation of the identification of an order $K=3$ plant. Similar results were obtained. The special feature here is the really distinct trajectory of the SML algorithm, which is caused by it being nonlinear in the parameters.

\begin{figure}[!t]
\centering
\includegraphics[clip = true, trim = 115 315 130 315, width=8.5cm]{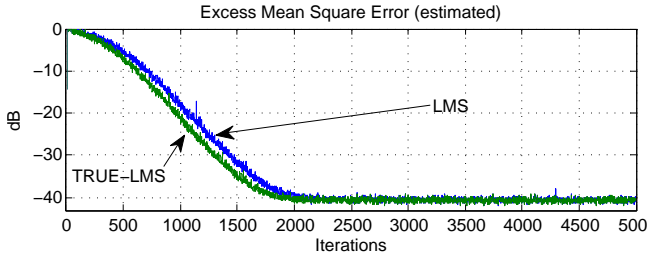}%
\caption{EMSE of the LMS and the TRUE-LMS.}
\label{fig:truelms}
\end{figure}

On the first $K =2$  plant, the LMS was simulated against the TRUE-LMS and the results are on Fig. \ref{fig:truelms}. The TRUE-LMS is shown as marginally faster while converging to the same EMSE as the LMS. Although the differences in the figure are minimal, the TRUE-LMS appeared to be a more stable algorithm. This can be explained by noting that the statistical parameters in its formulation are an average of samples, which helps remove outliers which otherwise would try to take the algorithm to a region of instability.

\begin{figure}[!b]
\centering
\subfloat[The Gaussian kernel with $\rho = 0.5$.]{
\includegraphics[clip = true, trim =120 280 115 270, width=8.5cm]{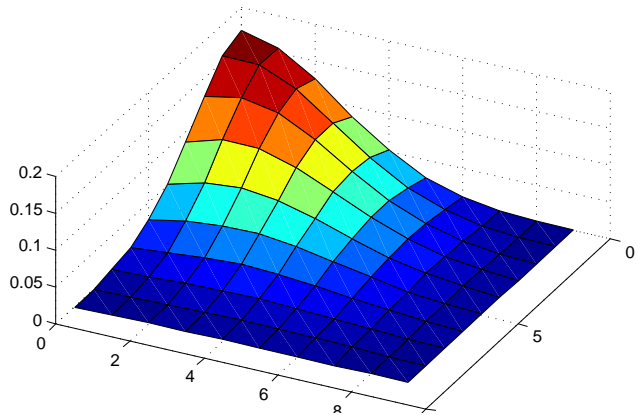}%
\label{gkernel}
}

\subfloat[A family of MSE curves for different values of $\rho$.]{\hspace{4mm}
\includegraphics[clip = true, trim = 115 245 110 250, width=8.5cm]{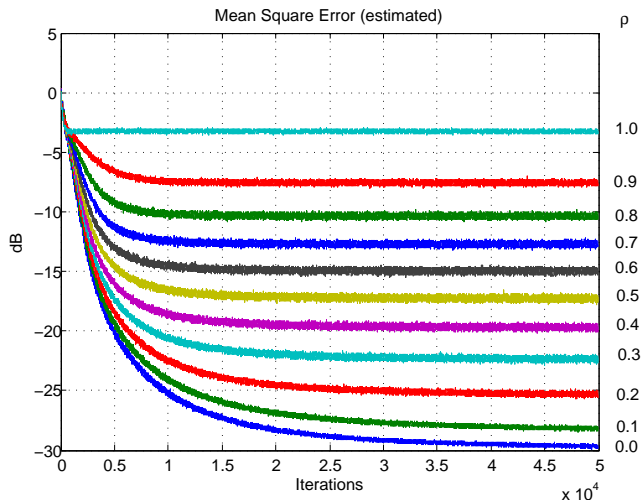}%
\label{mmse}
}
\caption{Tests on Gaussian plants for hypothesis of decomposability.}
\label{fig:gaussian}
\end{figure}

\subsection{Stability}

As previously addressed in subsection \ref{ssec:stable}, nonlinearities in the algorithm recursion may lead to non global stability. The following experiments use the same decomposable plant as the one used on Fig. \ref{ploti} and are intended to showcase the stability of the algorithms with various step-sizes. 

For this plant, the following parameter was computed:
\begin{equation}
\mu_0 = \lim_{i\to \infty}\frac{2}{3^K \tr \mathbb{E}\left[ (y_i^* y_i) \otimes (u_i^* u_i)\right]} \approx 0.1052.
\end{equation}

This experiment was run through a total $10.000$ realizations and the number of diverging realizations was counted. Table \ref{tb:stable} shows the results, for the LMS and the TRUE-LMS ($L = 4$ and $L = 8$), as a function of the step-size.

\begin{table}[t]
  \centering
\caption{Number if divergent realizations in the experiment.}
  \label{tb:stable}
  \begin{tabular}{c|ccccc}
  \hline
    & $0.5 \mu_0$ & $0.9 \mu_0$ & $\mu_0$ & $1.5 \mu_0$ & $2 \mu_0$ \\
    \hline
   LMS & $0$ & $0$ & $1$ & $50$ & $420$\\
   TRUE-LMS ($L = 4$) & $0$ & $0$ & $0$ & $12$ & $74$\\
   TRUE-LMS ($L = 8$) & $0$ & $0$ & $0$ & $0$ & $8$\\
  \hline
  \end{tabular}
\end{table}

It is possible to notice the stability of the algorithms for the steps lower than $\mu_0$. Moreover, they become more stable the greater the value of $L$, which was expected. Additionally, for the TRUE-LMS, the step-bound $\mu_0$ may be too conservative, but this depends in a complex way on the value of $L$.

\subsection{Adaptive algorithms as an approximation of the steepest descent}

\begin{figure}[!t]
\centering
\subfloat[Small $\mu$.]{
\includegraphics[clip = true, trim = 115 315 130 315, width=8.5cm]{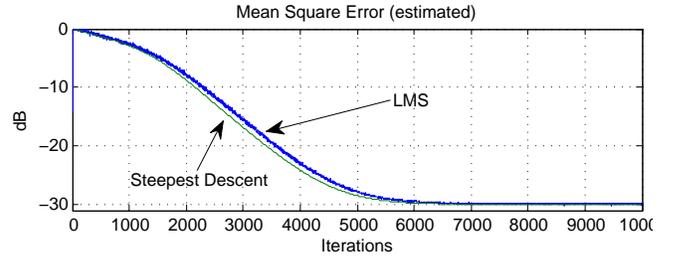}%
\label{fig:steeps}
}

\subfloat[Big $\mu$.]{
\includegraphics[clip = true, trim = 115 315 130 315, width=8.5cm]{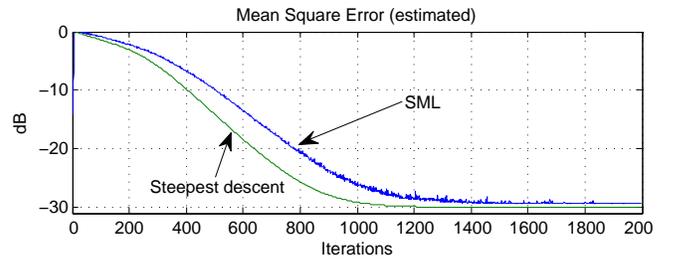}%
\label{fig:steepb}
}
\caption{Comparing the steepest descent algorithm with the LMS.}
\label{fig:lmsgrad}
\end{figure}

Continuing with the plant on Fig. \ref{ploti}, this section intends to show how the adaptive algorithms can be considered an approximation to the the steepest descent algorithm on \eqref{eq:steep}, as long as the step-size is small enough. Here the curves of the TRUE-LMS are not shown, as they would be almost superimposed on the ones of the LMS, as it can be seen on Fig. \ref{fig:truelms}. The curves of the adaptive algorithms form an ensemble average of $10.000$ realizations and are present Fig. \ref{fig:lmsgrad}.

When the step-size is small, as in Fig. \ref{fig:steeps}, the curves are almost superimposed. When it is increased, they start to take different trajectories. Another noticeable effect is the increase of the minimum MSE in the adaptive algorithm.

\subsection{A family of non-decomposable plants}
\label{ssec:howwell}

In this section the hypothesis of decomposability is studied. A Volterra kernel given by a bivariate normal probability function was chosen as the family of test plants. The reason for this is that the the correlation coefficient $\rho$ turns out to be a good measure of the decomposability of the plant, with $\rho = $ being perfect decomposability and $\rho = 1$ the worst-case scenario. In explicit terms, the plant is given by
\begin{equation}\small
	w_o(i, j) = \alpha \exp\left[-\frac{(i-1)^2 + (i-1)^2 +2\rho(i-1)(j-1)}{18 (1 - \rho^2)} \right],
\end{equation}
where $\alpha$ is a normalization parameter. A typical plant is shown on Fig. \ref{gkernel}.

The tests were made with $\sigma_v^2 = 10^{-3}$ and by varying the $\rho$ parameter between $0$ and $1$ by increments of $0.1$. The results of the tests are on Fig. \ref{mmse}. As predicted, the best results were with low values of $\rho$ and the worst one were with high values. This can be explained by thinking about the SVD. When $\rho \approx 0$, the singular value distribution of $w_o$ is concentrated in a single one. It is to the matrix associated to this singular value that the algorithm converges. Since this matrix contains most of the energy of the system, the minimum MSE is low. Conversely, with $\rho \approx 1$ the distribution of singular values of this matrix is more uniform. Even if the algorithm converges to the term with most energy, there will be a lot of energy remaining in the system, which corresponds to a higher minimum MSE.

\subsection{SML versus algorithms in the literature}
\label{ssec:literature}

The algorithm was tested against various others from the literature, with $K = 2$. The ones tested were:
\begin{itemize}
	\item The Power Filter (PF) \cite{pfilter} on \eqref{eq:simpvolt} with $D = 1$, which uses only the diagonal elements of the Volterra series. \textbf{22 coefficients.}
	\item The Simplified Volterra Filter (SV) \cite{svolt,svolt2}, the truncated diagonals model in \eqref{eq:simpvolt}, with $D = 3$.. \textbf{60 coefficients.}
	\item The Sparse Interpolated Volterra (IV) \cite{batista1,batista2}, which estimates only a few entries of the Volterra kernel and interpolates the others. \textbf{66 coefficients.}
	\item The full Volterra model (V) on \eqref{eq:volt}. \textbf{231 coefficients.}
	\item The SML-LMS algorithm on \eqref{eq:lms}. \textbf{42 coefficients.}
\end{itemize}

\begin{figure}[!t]
\centering
\subfloat[Representation of the smooth plant used in the simulation against other filters in the literature.]{
\includegraphics[clip = true, trim = 120 280 115 270, width=8.4cm]{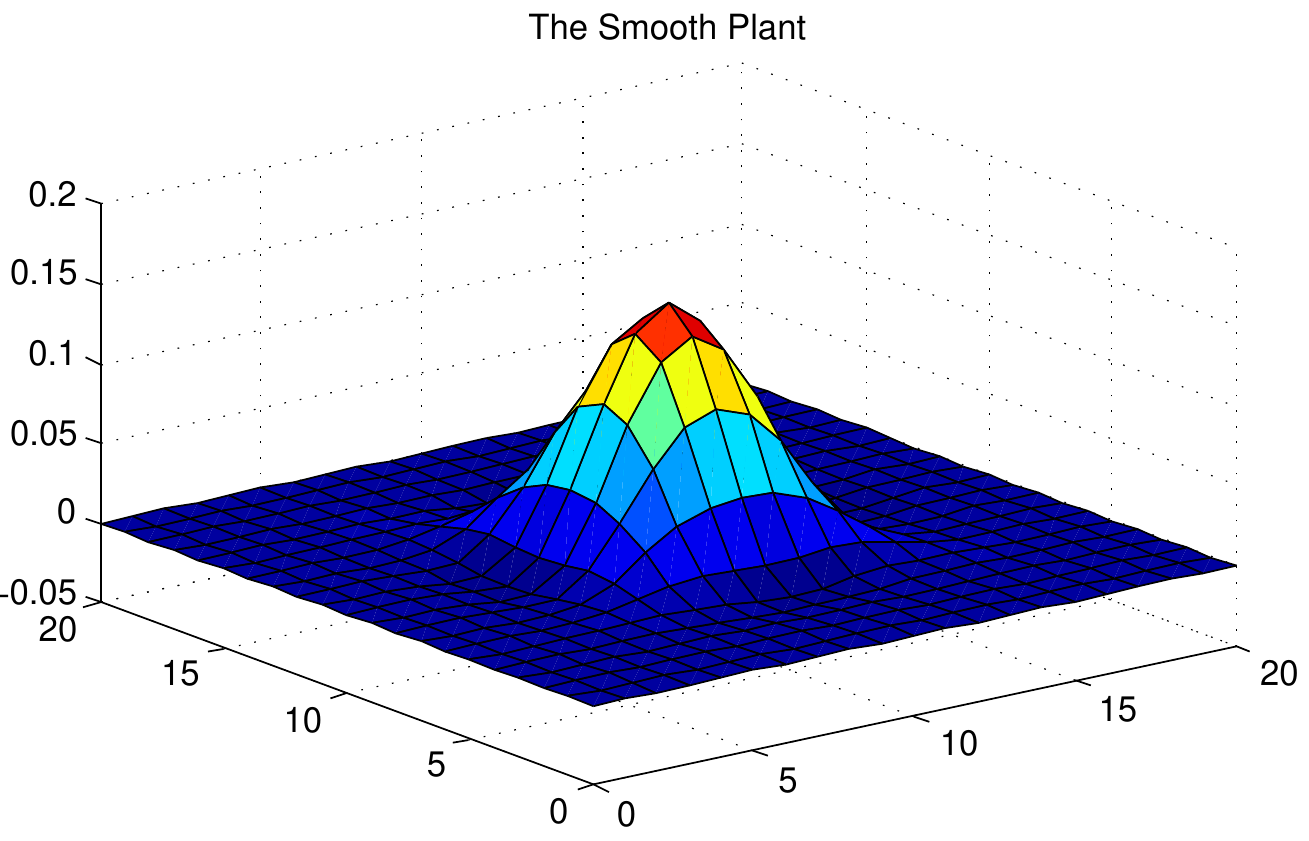}%
\label{fig:smooth}
}

\subfloat[Various filters identifying a smooth plant.]{
\includegraphics[clip = true, trim = 200 315 200 325, width=1.58in]{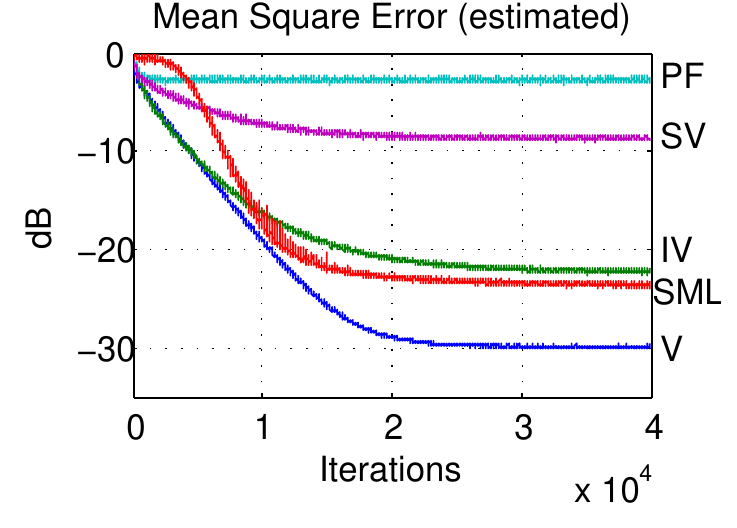}%
\label{plotv}
}\hspace{2mm}
\subfloat[Parallel cascade and SML.]{
\includegraphics[clip = true, trim = 200 315 200 325, width=1.58in]{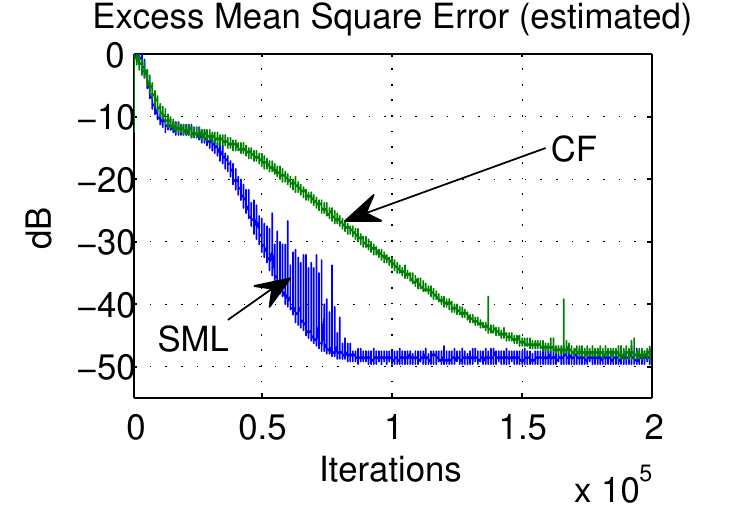}%
\label{plotvi}
}
\caption{Curves showcasing the performance of the SML algorithm against other algorithms in the literature.}
\label{fig:literature}
\end{figure}

The plant to be identified is a smooth one, as in \cite{batista2}. It is represented in \ref{fig:smooth}. Here the memory parameter was $M = 21$. The results are on Fig. \ref{plotv}. It shows the fast convergence and low minimum MSE of the SML filter, doing so with only 42 coefficients.

Fig. \ref{plotv} shows a simulation against the Parallel Cascade Filter \cite{cascade} (CF -- \textbf{65 coefficients}), a model that can be represented as a product of exactly two Volterra filters, and the SML (\textbf{20 coefficients}). They try to identify an $K = 3$ random decomposable plant. This experiment is here to showcase an structurally similar algorithm to the SML. They are both formed by products and are nonlinear in the parameters, but in this case the SML fares better. What is important to notice from this figure is that both algorithms have similar convergence paths, something that didn't happen with any of the others.

\section{Chaotic Behavior}
One last experiment concerns the emergence of chaotic behavior of the algorithm for certain values of the step-size. Assume a scalar model, with $M = 1$, and $K =2$. In this simple case, the algorithm will be adapting two parameters. Moreover, assume the model
\begin{equation}
d(i) = 100 u(i)^2,
\end{equation}
where $u(i) = 1$ is a constant signal. 

\begin{figure}[!t]
\centering
\includegraphics[width=8.4cm]{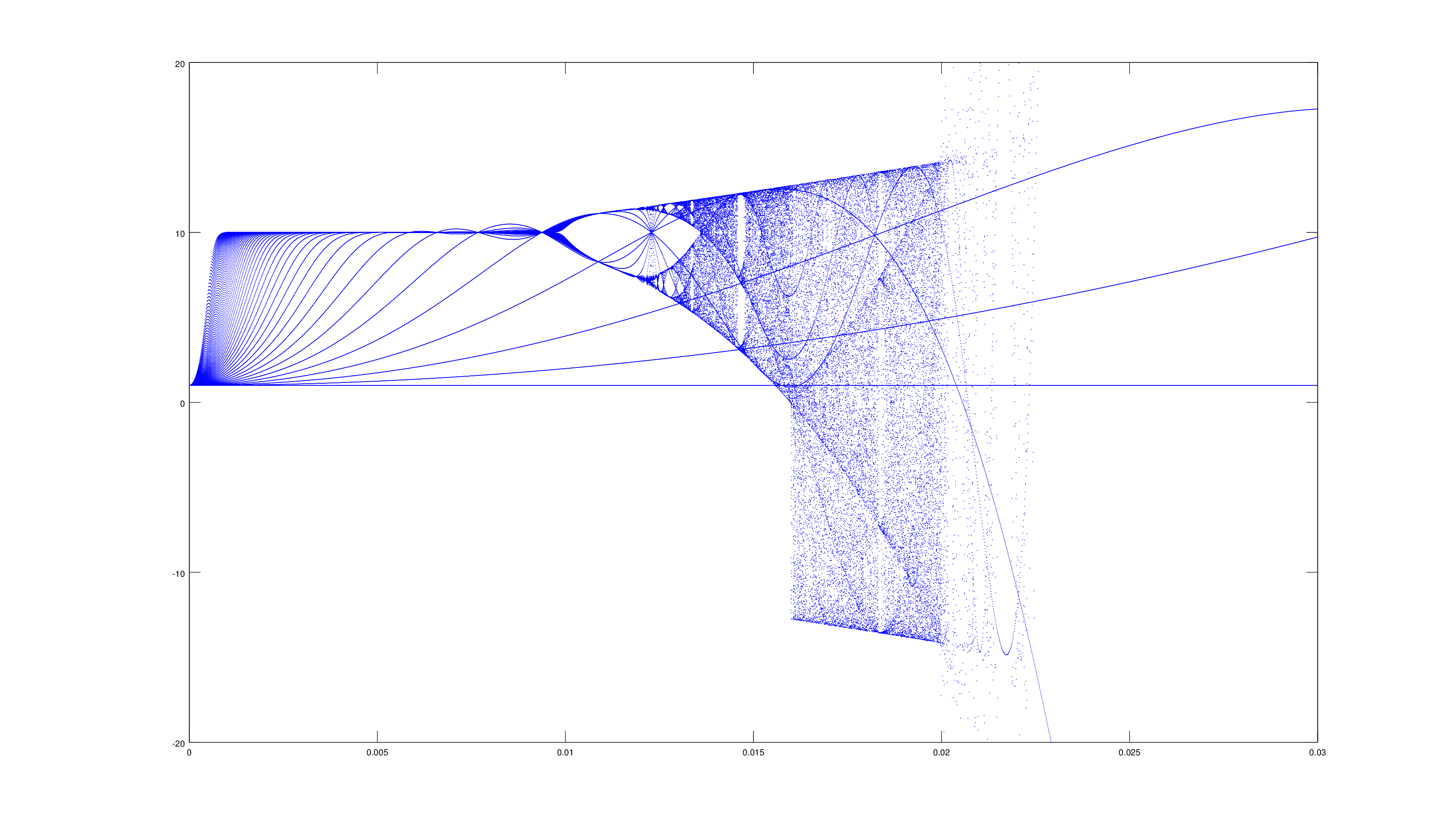}%
\caption{The bifurcation diagram of the system.}
\label{fig:chaos1}
\end{figure}

\begin{figure}[!t]
\centering
\includegraphics[width=8.4cm]{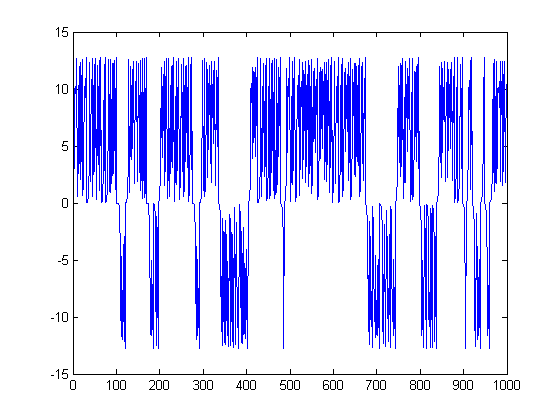}%
\caption{Trajectory of $w_1$ a $\mu\approx 0.016$.}
\label{fig:chaos2}
\end{figure}

The bifurcation diagram of the LMS algorithm, parametrized by $\mu$, is given by Fig. \ref{fig:chaos1}. it is possible to recognize a well-behaved convergence region in $0 < \mu < 0.01$, as predicted by step bounds. From there on, convergence starts to become oscillatory and starts to experience period doubling, until a point that is becomes fully chaotic. Starting from $0.016$, the parameter starts jumping from positive to negative regions. A typical trajectory in this region is presented in Fig. \ref{fig:chaos2}.

When for $\mu > 0.02$, the algorithm starts to diverge, but until $0.023$ this divergence seems chaotic. From there on it starts to appear monotonic, as expected.

Such behavior may be used in the way of improving convergence properties, such being part of a method to search for global minimums in multi-modal surfaces. This shall be a topic of investigation in further publications. 


\section{Conclusion}

This paper presents a low-complexity nonlinear adaptive filter based of the rank one approximation of the Volterra series. This is motivated both by the exponential reduction in complexity and in the well-posedeness of the problem of rank-one approximations.

The rank-one model can be represented as a product of linear FIR filters, which results in algorithms with complexity $O(KM)$, as opposed to the initial $O(M^K)$ of the full Volterra series. 

The description of this model is followed by the posing of an estimation problem. Its solution is given in the form of a steepest descent algorithm, which, as in classical adaptive filter theory, can make for the ideal model of an adaptive filter---they are approximations of the exact gradient algorithm. Indeed, simulations show that in the limit of small step sizes, the adaptive algorithms are expected to approximate really well the exact algorithm.

These adaptive filters derived as an rectangular window instantaneous approximation of the steepest descent. Depending on the number of samples on the window, two algorithms were derived: the LMS, with one sample, and the TRUE-LMS, which uses a general $L$ number of samples. Their trajectories can be verified to be roughly the same, but their properties differ in terms of stability.

Due to nonlinearities on the equations that implement the algorithms, it is expected that they do not achieve global asymptotic stability. An attempt at ameliorating this problem was done via a heuristic limitation of the algorithm. Other heuristic considerations were done in the choice of parameters for the filter. These considerations were later verified, albeit non-extensively, in simulations.

Other simulations involved testing of the algorithm in the identification of both rank-one and higher rank plants. It fares well in comparison with other algorithms when identifying many of these plants, specially those that can be considered approximately decomposable. The cases were the algorithms have poor performance were identified, for order $K = 2$ plants, as those whose Volterra kernels have evenly distributed singular values.

This paper finalizes with considerations on extensions. Some properties, like chaotic regions of the algorithm, may be studied further. Whether these properties either show to be detrimental to its performance or if reveal themselves to be of some application remains to be investigated. 

Other areas of extension is in the algorithms themselves. The solution to the estimation problem can be further studied by the used of Newton's method \cite{eusipco2016}. There is some work to be done in the ill-posedness of the general rank approximation and any development there could be applied in the development of general rank algorithms. Another extension is in the use of combinations of filters. For example, data-reuse-type algorithms, such as the TRUE-LMS, can have its performance improved by the use of incremental combinations. \cite{chamon} The same should apply in the algorithms developed here.


%

\appendices
\section{Norms and Inner Products of Tensors}
\label{sec:tensornorms}

Let $(V_1,  \langle\cdot,\cdot\rangle_1), (V_2,  \langle\cdot,\cdot\rangle_2), \dotsc, (V_K,  \langle\cdot,\cdot\rangle_K)$ be complex inner product spaces. The natural way of inducing an inner product on the tensor product $V_1\otimes V_2\otimes \dotsb\otimes V_K$ is by defining the inner product on the decomposable tensors as
\begin{align}
	\langle v_1\otimes v_2 \otimes \dotsb \otimes v_K, w_1\otimes w_2 \otimes \dotsb \otimes w_K\rangle \nonumber\\
	\triangleq \langle v_1, w_1 \rangle_1 \langle v_2, w_2 \rangle_2 \dotsm \langle v_K, w_K \rangle_K
\end{align}
and extending through sesquilinearity for general tensors, which are linear combinations of decomposable ones.

This inner product induces an $\ell_2$ norm in the usual way as $\|\mathcal{T}\| = \sqrt{\langle\mathcal{T}, \mathcal{T}\rangle}$. For decomposable tensors, this is expressed as
\begin{equation}
	\|v_1\otimes v_2 \otimes \dotsb \otimes v_K\| = \|v_1\|_1 \|v_2\|_2 \dotsm \|v_K\|_K,
\end{equation}
where each $\|\cdot\|_k$ is the respective induced $\ell_2$ norm. For general normed spaces the induced tensor norms are not as simply (or uniquely) defined, but reasonable ones should satisfy the cross norm identity on the decomposable tensors, which is also represented by the previous equation. On Banach spaces, an example of tensor norm is the projective norm, given by
\begin{align}
	\left\|\mathcal{T}\right\|_\pi = \inf\left\{\sum_i \|v_{1,i}\|_1 \|v_{2,i}\|_2 \dotsm \|v_{K,i}\|_K \colon\right.\nonumber\\ 
	\left.\sum_i v_{1,i}\otimes v_{2,i} \otimes \dotsb \otimes v_{K,i} = \mathcal{T}\right\},
\end{align}
that is, the smallest sum of norms of all the possible forms of writing $\mathcal{T}$ as a sum of decomposable terms. Another norm, called the injective norm, is defined in terms of linear functionals, but, regardless of the details, what is important is that these are all cross norms i.e. they separate on the decomposable tensors.

%
%

\ifCLASSOPTIONcaptionsoff
  \newpage
\fi



\bibliographystyle{IEEEtran}
\bibliography{IEEEfull,refs}
\end{document}